\documentclass[12pt]{article}
\usepackage{psfig,epsf}
\usepackage{epsfig,psfig,float}
\usepackage{a4wide,graphicx}
\usepackage{amssymb,amsbsy}
\usepackage{fancyhdr}
\usepackage{indentfirst}
\usepackage[notcite,notref]{}
\usepackage{cite}
\usepackage[spanish,english]{babel}

\def\be{\begin{eqnarray}}
\def\ee{\end{eqnarray}}

\begin{document}
\begin{flushright}
FTUV-00-0419 \\ 
IFIC-00-0419
\end{flushright}

\begin{center}
{\Large{\bf \mbox{\boldmath $\phi$} DECAY IN NUCLEI}}

\vspace{1cm}

{\large E. Oset$^1$ and
A. Ramos$^2$}

\vspace{0.4cm}

{\it $^1$ Departamento de F\'{\i}sica Te\'orica and IFIC,\\
\vspace{-0.2cm}
Centro Mixto Universidad de Valencia-CSIC, \\
\vspace{-0.2cm}
Institutos de Investigaci\'on de Paterna, Apdo. correos 2085,\\
\vspace{-0.2cm}
46071 Valencia, Spain}

{\it $^2$ Departament d'Estructura i Constituents 
de la Mat\`eria,
Universitat de Barcelona, \\
\vspace{-0.2cm}
Diagonal 647, 08028 Barcelona, Spain
}

\end{center}

\vspace{2cm}

\begin{abstract}
 
  We have studied the decay of the $\phi$ meson in nuclear matter by taking
into account the renormalization of the $K$ and $\bar{K}$ mesons in the
medium. We
separate the contribution of the total width into different reaction
channels, $K \bar{K}$ and $K Yh$ , the latter ones associated with the
$\phi N
\rightarrow K^+ Y $ reactions. We find a total width at normal nuclear
matter density of about 22 MeV, considerably larger than the free one.

\end{abstract}

\vspace{0.5 cm}

\noindent {\it PACS:} 13.15.-k, 13.75.Jz, 14.40.Ev, 25.75.-q

\noindent {\it Keywords:} $\phi$ decay in nuclei, Chiral Lagrangian,
${\bar K}$-nucleus interaction
\newpage

\section{Introduction}

The renormalization of the hadron properties in a nuclear medium is the
object of continuous attentions. Particularly, the properties of the
$\rho$
meson in nuclei have been thoroughly studied
\cite{beng,chanfray,johan,klinuc} and many experiments
are devoted to observe these modifications \cite{hades,ceres}.
Comparatively, the
$\phi$ meson has received little attention but the studies of
Refs.~\cite{klinuc,klilett} predict a very small shift, if any,
and a
substantially increased width.  Yet, the study of the $\phi$ width in a
nuclear medium is a very interesting question since it should offer
information on the renormalization of the kaon properties in
a medium, a subject itself which attracts much interest
\cite{koch,wolfram,lutz,angels} and which is related to the possible
existence of kaon condensates in neutron stars \cite{kaplan}. 
  
  The many body problem of the interaction of the $\bar{K}$ 
with a nuclear medium is a
subtle one.  The low density theorem  leads to a
repulsive $\bar{K}$ selfenergy, but analysis of kaonic atoms demands an
attractive one \cite{batty,gal,nieves}. A step forward in the understanding
of this peculiar feature is given in Refs.~\cite{koch,wolfram} where it
is shown
that the Pauli blocking of intermediate states in the $\bar{K} N$
interaction leads to a shift of the $\Lambda(1405)$ resonance (which lies
just below the  $\bar{K} N$ threshold) to higher energies. As a consequence
the $\bar{K} N$ interaction becomes attractive. On the other hand,
the
attraction felt
by the kaons has an opposite effect because it brings the resonance
back to
lower energies. This fact stimulated a selfconsistent calculation in
Ref.~\cite{lutz} which showed that the position of the $\Lambda(1405)$
resonance
was not changed in the medium but altogether 
the $\bar{K}$ still felt an attraction and the $\Lambda(1405)$
resonance broadened considerably. The
calculations of Ref.~\cite{angels} introduced in addition the
renormalization of
the pions and the baryons in the intermediate states, and found again that
the position of the resonance barely changed when the $\bar{K}$  selfenergy
was calculated selfconsistently while the width became even larger. The
${\bar K}$-nucleus potential of Ref.~\cite{angels} has been used
in the study of Ref.~\cite{satoru} where it was
found to be compatible with the present data of kaonic atoms. 

The important changes found in the kaon selfenergy in the medium, which
is the
key ingredient here, advise a reevaluation of the $\phi$ selfenergy and
this is the purpose of the present work.
The framework for the evaluation of the $\phi$ selfenergy in a nuclear
medium was developed in Ref.~\cite{klinuc,klilett} and we will adhere
to it,
yet following a different technical approach. 

\section{\mbox{\boldmath $\phi$} decay in the nuclear medium}

  We will use here the gauge vector representation of the vector
field
$\phi$. The alternative tensor formulation of Ref.~\cite{ecker} is
equivalent to
the gauge vector one but makes the chiral counting easier. Yet, in the
present problem one needs Lagrangians for the coupling of the mesons to
baryons which are available in the vector representation but not in the
tensor one and hence we shall adhere to the vector representation as done in
Refs.~\cite{klinuc,klilett}. 

  In free space the $\phi$ meson decays into $K \bar{K}$ and $3
\pi$. The latter
channel is OZI forbidden and in spite of the large phase space only accounts
for 15\% of the  total width. The coupling of the $\phi$ to 
  $K \bar{K}$ is large and it is only the reduced phase space what
makes the
$\phi$ width small. However, we shall see that thanks to the $\bar{K}$
related
channels in the medium, the $\phi$ width becomes of the order of 22 MeV at
normal nuclear matter density ($\rho_0=0.17$ fm$^{-3}$) and the $3 \pi$
channel represents only 3\% of
this width. Thus, we shall neglet it in our study.

   The Lagrangian that describes the coupling of
 the $\phi$ meson to kaons is given by
\begin{equation}
{\cal L}=-i g_\phi \phi_\mu (K^-\partial_\mu K^+ - K^+\partial_\mu K^-
+ {\bar K}^0\partial_\mu K^0 - K^0\partial_\mu {\bar K}^0 ) \ ,
\label{eq:lagran}
\end{equation}
which provides the following $\phi K^+ K^-$ vertex 
(matrix element of $ i{\cal L}$)
\begin{equation}
V_{\phi K^+ K^-} = - i g_\phi \epsilon_\mu (p^\mu-p^{\prime\,\mu}) \ ,
\end{equation}
with $p,p^\prime$ the momenta of the $K^+,K^-$ mesons,
respectively. An analogous vertex in found for
$K^0 {\bar K}^0$.

\begin{figure}[htb]
     \centering
      \epsfxsize = 6cm
      \epsfbox{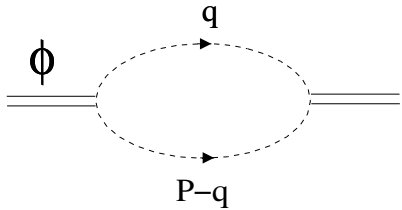}
      \caption{\small Selfenergy diagram contributing to the
decay of the $\phi$ into $K^+K^-$ and $K^0\bar{K}^0$ pairs.}
        \label{fig:phifig1}
\end{figure}

   We shall work in a frame where the $\phi$ is at rest and in a gauge where
$\epsilon^0=0$. If we consider the $\phi$ selfenergy diagram of
Fig.~\ref{fig:phifig1}
with
$K^+ K^-$ as intermediate states we obtain 
\begin{equation}
\Pi_\phi = i \int \frac{d^4 q}{(2\pi)^4} \,\frac{1}{q^2-m_K^2
+i\varepsilon}\, \frac{1}{(P-q)^2-m_K^2+i\varepsilon}\, g_\phi^2 \,
\epsilon_i
\epsilon_j 2 q_i 2 q_j \ ,
\end{equation}
where $P$ is the fourmomentum of the $\phi$.  The imaginary part of
$\Pi$ is
readily evaluated using Cutkosky rules
\begin{eqnarray}
\Pi_\phi &\to& 2 i {\rm Im\,} \Pi_\phi \nonumber \\
D(q) &\to& 2 i \theta(q^0) {\rm Im\,} D(q) \\
D(P-q) &\to& 2 i \theta(P^0-q^0) {\rm Im\,} D(P-q) \ ,
\end{eqnarray}
where $D(q)$ is the kaon propagator. Then, using the relationship
\begin{equation}
\Gamma_\phi = - \frac{{\rm Im}\, \Pi_\phi}{M_\phi} \ ,
\end{equation}
with $M_\phi$ the $\phi$ mass, we obtain the free width of the $\phi$ for
$K^+ K^-$ decay 
\begin{equation}
\Gamma_\phi = \frac{2}{3}\,\frac{1}{4\pi}\, g_\phi^2
\, \frac{p^3}{M_\phi^2} \ ,
\label{eq:free}
\end{equation}
which requires
the
value $g_\phi=4.57$ to obtain agreement with the experimental value of
2.18 MeV 
\cite{pdg}.

 In the nuclear medium the kaon acquires a selfenergy,
$\Pi_K(q^0,\vec{q},\rho)$,
and the
kaon propagator reads
\begin{equation}
D(q,\rho) =
\frac{1}{q^{0\,2}-\vec{q}\,^2-m_K^2-\Pi_K(q^0,\vec{q},\rho)} \ ,
\label{eq:prop}
\end{equation}
which can also be cast in terms of the Lehmann representation as
\begin{equation}
D(q,\rho)=\int_0^\infty d\omega \, 2\omega\,
\frac{S_K(\omega,\vec{q},\rho)}{q^{0\, 2}-\omega^2+i\varepsilon} \ ,
\label{eq:lehmann}
\end{equation}
where $S_K$ is the kaon spectral function
\begin{equation}
S_K(p,\rho)=-\frac{1}{\pi} {\rm Im\,} D(p,\rho) \ .
\label{eq:spec}
\end{equation}

At this point it is worth differentiating between the $\bar{K}\equiv
(\bar{K}^0, -K^-)$ and the
$K\equiv (K^+,K^0)$ meson doublets.
 
The $K^+ N$ and $K^0 N$ interactions are smooth, there
are no resonances with strangeness
$S=1$
and the selfenergy of the $K^+$ meson at small energies is
well
aproximated by the
$t \rho$ approximation, which from Refs. \cite{siegel,knangels} is
given by
\begin{equation}
\Pi_{K^+}=\frac{1}{2}\,(t_{K^+ p}+t_{K^+ n})\,\rho_0
\, \frac{\rho}{\rho_0} =
0.13\, m_K^2 \,\frac{\rho}{\rho_0}   \ . 
\label{eq:selfplus}
\end{equation}
An identical expression is found for the $K^0$ meson.
 
The $\bar{K}$ case is far more subtle. As we pointed out, 
the $\bar{K} N$ interaction at low energies is dominated by the
$\Lambda(1405)$ resonance which
appears just below the $\bar{K} N$ threshold.  Any realistic approach to the
problem has do deal properly with this resonance, which is actually one of
the resonances which meets with more problems within
quark models of
baryons. Not surprisingly, the resonance is closely tied to the
interaction
of the $\bar{K} N$ system with related channels and is generated within
unitary coupled-channel schemes
\cite{dalitz,martin,siegel2}. An
important step
forward has been given with the use of chiral Lagrangians within a  
coupled channel unitary approach \cite{siegel,knangels}, by means of
which one is able to reproduce the $\Lambda(1405)$ resonance and the cross
sections for $K^-p$ scattering to the different coupled channels. This
succesful scheme
is most appropriate to undertake the study of the kaon selfenergy in a
nuclear medium and this was done in Refs.~\cite{wolfram,lutz,angels}.
As quoted above, 
with respect to the work of Ref.~\cite{wolfram}, 
the work of Ref.~\cite{lutz} implemented
the selfenergy of the kaon in the calculation in a selfconsistent
way, with the important finding that the position of the $\Lambda(1405)$
resonance was barely moved and the $K^-$ experienced a moderate
attraction.
The work of Ref.~\cite{angels} improves on the one of Ref.~\cite{lutz}
by including
the selfenergy of the pions and the baryons in the intermediate states, thus
opening more channels for reactions of the $K^-$ in the medium, 
consequently increasing the width of the $\Lambda(1405)$ resonance 
and widening considerably the spectral function of the $K^-$
(see also Ref.~\cite{report} for a recent review on this and related
issues). Clearly, the spread of the kaon strength
necessarily
should have repercussion on the $\phi$ decay in the medium since it
automatically enlarges the phase space for the $\phi$ decay into the kaon
related channels. 

The low energy coupled channel equations studied in
Refs.~\cite{siegel,knangels} deal only with 
the s-wave part of the ${\bar K}N$ interaction. 
A recent study of the p-wave $K^- N $
interaction
at low energies within the context of chiral Lagrangians has been done in
Ref.~\cite{caro}. 
Although the energies of the kaons in 
the $\phi$ decay
are not large and hence the s-wave selfenergy plays the major role in the
process, we nevertheless take the p-wave into account by including
the
excitation of the $\bar{K}$ into $\Lambda
h, \Sigma h $ and $\Sigma^* h$ states, as done
in Refs.~\cite{klinuc,klilett},  
where $\Sigma^*$ is the $\Sigma(1385)$ resonance \cite{pdg}. 
The p-wave selfenergy plays a role in the decay of the $\phi$
into $K^+ Yh$ states, while the s-wave selfenergy contributes
essentially 
to the 
decay into $K^+ \Sigma\pi h$ states.

In the present work we use the s-wave $K^-$ selfenergy from
Ref.~\cite{angels}
and implement the p-wave selfenergy due to $\Lambda h, \Sigma h $
states, as also done
in Ref.~\cite{angels}. As just mentioned, we also include here 
the coupling of the
$\bar{K}$ into $\Sigma^* h$
excitations.
Its addition in the kaon loops in the selfconsistent
calculation of the s-wave $K^-$ selfenergy barely modifies the results of
Ref.~\cite{angels}. We will see, however, that as an extra contribution
to the p-wave selfenergy
it induces
an increase in the $\phi$ width, since it
incorporates the 
$\phi \rightarrow K^+ \Sigma^* h$ decay channel. 

  In order to evaluate the $\phi$ selfenergy in the medium we write the
$K^+$ propagator as in eq.~(\ref{eq:prop}) using explicitly the
$K^+$ selfenergy of
eq.~(\ref{eq:selfplus}). For the $K^-$ propagator we
use instead the
Lehmann representation of eq.~(\ref{eq:lehmann})  and thus, following
the same steps that
led us to eq.~(\ref{eq:free}), we obtain a $\phi$ width in the medium
given by
\begin{eqnarray}
\Gamma_\phi(P^0,\rho) &=&
\frac{1}{M_\phi} \int_0^{\omega_{\rm max}^{(+)}}
d\omega\, S_{K^-}(\omega,q^{(+)},\rho) \, \frac{1}{3\pi}\,  g_\phi^2 \,
q^{(+)\, 3} \nonumber \\
& + &
\frac{1}{M_\phi} \int_0^{\omega_{\rm max}^{(0)}}
d\omega\, S_{\bar{K}^0}(\omega,q^{(0)},\rho) \, \frac{1}{3\pi}\,  g_\phi^2 \,
q^{(0)\, 3} \ ,
\label{eq:medium}
\end{eqnarray}
where
\begin{eqnarray}
q^{(+)}= \sqrt{ (P^0-\omega)^2 - \Pi_{K^+}-m_{K^+}^2 } \ , & 
\omega_{\rm max}^{(+)} = P^0 - \sqrt{m_{K^+}^2+\Pi_{K^+}} \nonumber
\\
q^{(0)}= \sqrt{ (P^0-\omega)^2 - \Pi_{K^0}-m_{K^0}^2 }\ , &
\omega_{\rm max}^{(0)} = P^0 - \sqrt{m_{K^0}^2+\Pi_{K^0}} \ .
\end{eqnarray}

  The vertex $\bar{K} N Y$ for an incoming kaon of momentum $\vec{k}$
is given by
\begin{equation}
V_{\bar{K}NY} = \widetilde{V}_{\bar{K}NY}\, \vec{\sigma}\cdot\vec{k} =
\left[ \alpha \frac{D+F}{2f} + \beta \frac{D-F}{2f} \right]
\vec{\sigma}\cdot\vec{k} \ ,
\label{eq:couplings}
\end{equation}
where $\alpha$ and $\beta$ are given in Table \ref{table1}
and $D+F=g_A=1.257$, $D-F=0.33$. We take 
$f=1.15 f_\pi$ with $f_\pi=93$ MeV, a value which
lies in between the pion and kaon weak decay constants and that was
chosen in the model of Ref.~\cite{knangels} to optimize the position of 
the $\Lambda(1405)$ resonance.

\begin{table}[ht]
\centering
\caption{\small Coefficients for the $\bar{K}NY$ couplings of eq.
(\ref{eq:couplings})}
\vspace{0.5cm}
\begin{tabular}{c|cccccc}
        & $K^-p\to\Lambda$ & $K^-p\to \Sigma^0$ & $K^- n\to \Sigma^-$
&
$\bar{K}^0 n \to \Lambda$ & $\bar{K}^0 n \to \Sigma^0$ &
$\bar{K}^0 p \to \Sigma^+$ \\
        \hline
$\alpha$ & $-\frac{2}{\sqrt{3}}$ & 0 & 0 & $-\frac{2}{\sqrt{3}}$ & 0
&
0 \\
$\beta$ & $\frac{1}{\sqrt{3}}$ & 1 & $\sqrt{2}$ &
$\frac{1}{\sqrt{3}}$
& $-1$ & $\sqrt{2}$ 
\end{tabular}
\label{table1}
\end{table}

Similarly, the $\bar{K} N \Sigma^*$ vertex is given by
\begin{equation}
V_{\bar{K}N\Sigma^*} = \widetilde{V}_{\bar{K}N\Sigma^*}\,
\vec{S}\,^\dagger\cdot\vec{k} =
\frac{g_{\scriptscriptstyle{\Sigma^*}}}{2M}\,
 A \,
\vec{S}\,^\dagger\cdot\vec{k}    \ ,
\label{eq:sigma}
\end{equation}
where $\vec{S}\,^\dagger$ is the spin transition operator from spin 1/2
to spin 3/2.
The coefficient A is given in Table \ref{table2} and the coupling
$g_{\scriptscriptstyle{\Sigma^*}}/2M$ is
evaluated here by first using the SU(6) quark model to relate the $\pi
N N$ coupling
to the $\pi N \Delta$ one and then
using SU(3) symmetry to
relate the $\pi N \Delta$
coupling to the $\bar{K} N \Sigma^*$ one, 
since the $\Sigma^*$
belongs to the SU(3) decouplet of the
$\Delta$. 
We obtain:
\begin{equation}
\frac{g_{\scriptscriptstyle{\Sigma^*}}}{2M} = 
\frac{2\sqrt{6}}{5} \frac{D + F}{2f} \ .
\end{equation}
 There is hence a small diversion of our approach
here with
respect
to the one of Ref.~\cite{klinuc}, where SU(3) arguments are used and,
consequently, a different coupling was obtained.
We
note that
with the coupling used in the present work
one gets a good reproduction of the data for the decay of the
$\Sigma^*$
into $\Lambda \pi$ and $\Sigma \pi$ final statesx, when the
the value of $f$ is taken as $f_\pi$.

\begin{table}[ht]
\centering
\caption{\small Coefficient for the $\bar{K}N\Sigma^*$ couplings of
eq.~(\ref{eq:sigma})}
\vspace{0.5cm}
\begin{tabular}{c|cccc}
        & $K^-p\to\Sigma^{*0}$ & $K^- n\to \Sigma^{* -}$ & 
$\bar{K}^0 p \to \Sigma^{* +}$ & $\bar{K}^0 n \to \Sigma^{* 0}$ \\
        \hline
$A$ & $-\frac{1}{\sqrt{2}}$ & $-1$ & $-1$ & $\frac{1}{\sqrt{2}}$ 
\end{tabular}
\label{table2}
\end{table}

  The p-wave $\bar{K}$ selfenergy in symmetric nuclear matter can then be
written as
\begin{eqnarray}
\Pi^{(p)}_{\bar{K}}(q^0,\vec{q},\rho) &=&
\widetilde{\Pi}_{\bar{K}}(q,\rho) \vec{q}\,^2 \nonumber \\
&=& 
\frac{1}{2} \widetilde{V}^2_{K^- p \Lambda} f_\Lambda^2 {\vec q}\,^2
U_\Lambda(q^0,\vec{q},\rho) \nonumber \\
&+& \frac{3}{2} \widetilde{V}^2_{K^- p \Sigma^0}
 f_\Sigma^2 {\vec q}\,^2 U_\Sigma(q^0,\vec{q},\rho)  \\
&+& \frac{1}{2} \widetilde{V}^2_{K^- p \Sigma^{*0}}
 f_{\Sigma^*}^2 {\vec q}\,^2 U_{\Sigma^*}(q^0,\vec{q},\rho) \ , \nonumber
\label{eq:selfkap}
\end{eqnarray}
where the Lindhard function $U_Y(q)$ $(Y=\Lambda,\Sigma$ or $\Sigma^*$) is 
given by
\begin{eqnarray}
{\rm Re}\, U_Y(q^0,\vec{q},\rho) &=&
\frac{3}{2}\rho \frac{M_Y}{q p_F} \left\{ z + \frac{1}{2}(1-
z^2) \ln \frac{\mid z + 1\mid}{\mid z - 1 \mid} \right\}
\nonumber \\
{\rm Im}\, U_Y(q^0,\vec{q},\rho) &=& -\pi \frac{3}{4} \rho
\frac{M_Y}{q p_F} \left\{ (1-z^2) \theta(1-\mid z \mid)
\right\} \\
z &=&\left( q^0 - \frac{q^2}{2M_Y} - (M_Y-M) \right) \frac{M_Y}{q
p_F} \nonumber \ ,
\end{eqnarray}
with $\rho$ the nuclear matter density, $p_F$ the Fermi momentum
and $f_\Lambda=(1-q^0/2M_\Lambda)$, $f_\Sigma=(1-q^0/2M_\Sigma)$,
$f_{\Sigma^*}=(1-q^0/M_{\Sigma^*})$
relativistic recoil vertex corrections \cite{angels}.

As in Ref.~\cite{klilett}, we also incorporate a form-factor at 
the kaon-baryon vertices of dipole type, $[\Lambda^2/(\Lambda^2-q^2)]^2$,
with $\Lambda=1.05$ GeV.

\begin{figure}[htb]
     \centering
      \epsfxsize = 14cm
      \epsfbox{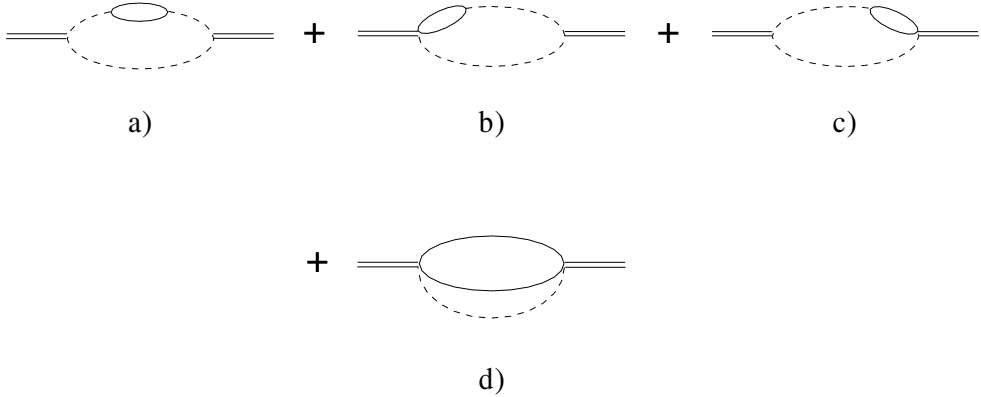}
      \caption{\small Selfenergy diagrams at first order in the nuclear
density contributing to the decay of the $\phi$ meson in the medium.}
        \label{fig:phifig2}
\end{figure}

Inserting
the p-wave $K^-$ selfenergy at the lowest order in the nuclear
density accounts for the contribution to the in medium $\phi$
selfenergy depicted in diagram a) of Fig.~\ref{fig:phifig2}.
However,
it was already noticed in Refs.~\cite{beng,chanfray} that, when dealing
with gauge vector mesons, there are other vertex corrections associated
to these
diagrams which are required by the gauge invariance of the model.  These
diagrams are depicted in Figs.~\ref{fig:phifig2}b), c) and d).
These terms appear from a
contact $\phi K N Y$ term, similar to the Kroll Ruderman term for
photons, and
they can be evaluated systematically by substituting $e Q A_\mu$ by
$g V_\mu/2$ in the chiral Lagrangians, as done in Ref.~\cite{klinuc},
with
$g=\displaystyle\frac{\sqrt{2}}{3} g_\phi$, where an extra minus sign is already
incorporated in
the
Lagrangian of eq.~(\ref{eq:lagran}).  We thus obtain the following vertex
function for these contact
terms
\begin{eqnarray}
V_{\phi KNY}&=&g_\phi \, \widetilde{V}_{\bar{K}NY} \, \vec{\sigma}\cdot
\vec{\epsilon}\,(\phi) ~ ; ~~~~~~~ Y=\Lambda,\Sigma \nonumber \\
V_{\phi KN\Sigma^*}&=&g_\phi \, \widetilde{V}_{\bar{K}N\Sigma^*} \,
\vec{S}\,^\dagger\cdot
\vec{\epsilon}\,(\phi)  \ .
\end{eqnarray}

  The contribution of the vertex corrections can be easily evaluated. We can
see that
the addition of all the one-baryon loop diagrams in Fig.~\ref{fig:phifig2}
just replaces the
contribution of
the one in Fig.~\ref{fig:phifig2}a),  $D(q)\widetilde{\Pi}(q)\vec{q}\,^2 D(q)$,
by 
\begin{equation}
D(q)\widetilde{\Pi}(q)\vec{q}\,^2 D(q) + \frac{1}{2} \widetilde{\Pi}(q)
D(q) + \frac{1}{2}
D(q)\widetilde{\Pi}(q) + \frac{3}{4} \frac{\widetilde{\Pi}(q)}{\vec{q}\,^2} \ .
\label{eq:all}
\end{equation}

  One simplifying step forward is possible at this point since we are
interested in the imaginary part of the selfenergy. By means of Cutkosky
rules one knows that the contributions to the imaginary part are obtained
when placing either
the $K^+ K^- $ or the $K^+ Yh$ of the intermediate states on shell. In both
cases the $K^+$ appears on shell, which means that we can substitute $P^0
- q^0$ by $\omega(q)$, hence $q^0= P^0-\omega(q)$, in eq.~(\ref{eq:all}).
It is then easy to see that the sum of all diagrams in
Fig.~\ref{fig:phifig2} is  
equivalent to considering only diagram a) but with the following
substitution for
the kaon p-wave selfenergy 
\begin{equation}
\widetilde{\Pi}(q)\vec{q}\,^2 \to \widetilde{\Pi}(q) [(P^0-\omega(q))^2-m_K^2]
\left\{ 1 + \frac{3}{4} \frac{[(P^0-\omega(q))^2-\vec{q}\,^2-m_K^2]^2}
{\vec{q}\,^2[(P^0-\omega(q))^2-m_K^2]}\right\} \ .
\label{eq:vertex}
\end{equation}

   The arguments given above apply to just the p-wave diagrams
at lowest order in the density, i.e. those displayed in
Fig.~\ref{fig:phifig2}.
In the actual calculation, however, we simply replace the p-wave
selfenergy by means of eq.~(\ref{eq:vertex}) in the $K^-$ propagator which is
used to
evaluate the $\phi$ decay width in the medium, hence generalizing the
correction of eq.~(\ref{eq:vertex}) to higher orders of the density. This
approximation is
well justified because when we pick up the imaginary part from the two kaon
cut the momenta involved are small and the p-wave plays a minor role. On the
other hand, the p-wave part becomes relevant when we pick up the $K^+ Yh$
excitation, in which case there is plenty of phase space available and the
momenta involved are large. In this case,  however, the $\bar{K}$ 
is far off shell and an additional particle-hole insertion
in the ${\bar K}$ selfenergy,
i.e. a higher order in the density, 
does not modify much the ${\bar K}$ propagator. 

The use of the technique
explained above simplifies the analytical structure of the integrand of the
$\phi$ selfenergy and allows one to use the same formalism that leads to eq.
(\ref{eq:medium}),
where there are only p(and s)-wave selfenergy insertions in the propagators 
and no
vertex corrections. 

\section{Results}

\begin{figure}[htb]
     \centering
      \epsfxsize = 10cm
      \epsfbox{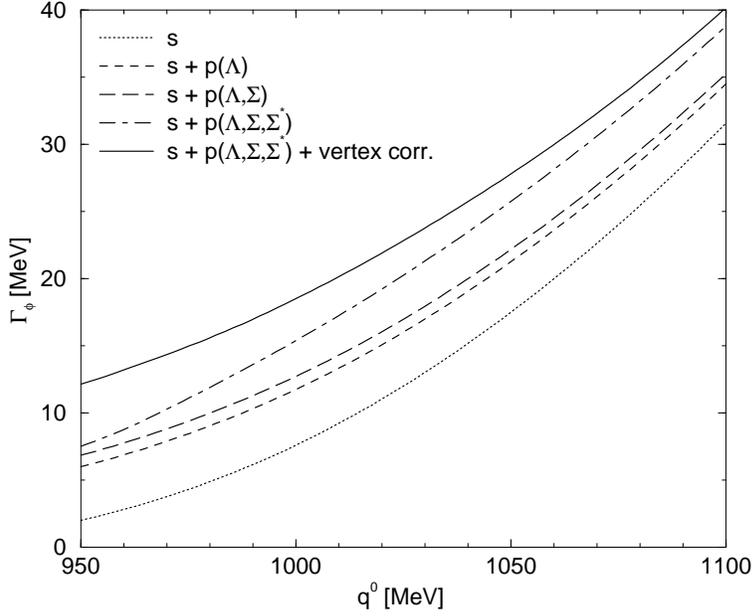}
      \caption{\small Width of the $\phi$ meson in nuclear matter at
density $\rho=\rho_0$ as a function of the $\phi$ energy.}
        \label{fig:phifig3}
\end{figure}

  In Fig.~\ref{fig:phifig3}  we show the different contributions to the
$\phi$ width as a
function of the $\phi$ energy. The dotted line represents the contribution
to the $\phi$ width when the $\bar{K}$ is dressed with only the s-wave
 selfenergy (the $K$ is dressed with the repulsive selfenergy of eq.
(\ref{eq:selfplus}) in
all these curves). We can already see an appreciable increase of the width
with respect to the free value of about 4 MeV. This is due on the one
hand
to the fact that the $\bar{K}$ feels an attraction of bigger strength than
the repulsion felt by the $K$, hence the phase space for $K \bar{K}$
decay increases.  On the other hand, the $\bar{K}$ in the nucleus decays
in
s-wave into the channels $\pi \Sigma h$ and $\pi \Lambda h$ and, therefore,
the
incorporation of the $\bar{K}$ s-wave selfenergy in the evaluation of the 
$\phi$ selfenergy
automatically accounts for the $\phi \to K \pi \Sigma h$ and  $\phi \to K
\pi \Lambda h$ channels or, conversely, the nucleon induced $\phi$ decay
reactions $\phi N \to K \pi
\Sigma $ and $\phi N \to K \pi \Lambda$.

  A next step in the figure incorporates the contribution of
decay channels
related to the p-wave $\bar{K}$ selfenergy through $Yh$ excitations, with
$Y=\Lambda$ (short dashed curve), $Y=\Lambda, \Sigma$ (long-dashed curve)
and $Y=\Lambda,\Sigma,\Sigma^*$ (dot-dashed curve). These correspond to
the reactions $\phi N \to K \Lambda, K \Sigma, K \Sigma^*$. We can see
that the largest contribution corresponds to the $\Lambda h$ excitation,
as already noted in Ref.~\cite{klilett}, 
since it has a larger phase space and in addition the $\bar{K}N\Lambda$
coupling
involves the $D+F$ combination, while the $\bar{K}N\Sigma$ one involves
the $D-F$ combination, which is about a factor four smaller. The
contribution of the $\Sigma^*$  is comparatively larger than the
$\Sigma$
one, due to the larger coupling, but still smaller than that of the
$\Lambda$ due to the reduced phase space.

Finally, the solid line in Fig.~\ref{fig:phifig3} shows the results obtained
when the vertex corrections are also incorporated.
They increase the width of the $\phi$ by an additional
3 MeV at the $\phi$ mass, $M_\phi=1019.413$ MeV.

The final width of the $\phi$ meson at the $\phi$ mass in symmetric nuclear
matter of density
$\rho=\rho_0$ amounts to 22 MeV.
All the different decay channels contributing to the final width increase
smoothly with energy as a
consequence of the increasingly larger available phase space.

Although not shown in the figure, the effect of the relativistic recoil
corrections
(factors $f_Y$ in eq.~(\ref{eq:selfkap})) amounts to a reduction of about 5
MeV, also at
the $\phi$ mass.   If the dipole form factor is omitted the
$\phi$ width raises to about 30 MeV. If, on the other hand, 
one uses a monopole factor of the type $\Lambda^2/(\Lambda^2-q^2)$,
with $\Lambda=1.3$ GeV, as commonly used in phenomenological studies 
of the hyperon-nucleon interaction, the $\phi$ width turns out to be
around 26 MeV.   

\begin{figure}[htb]
     \centering
      \epsfxsize = 10cm
      \epsfbox{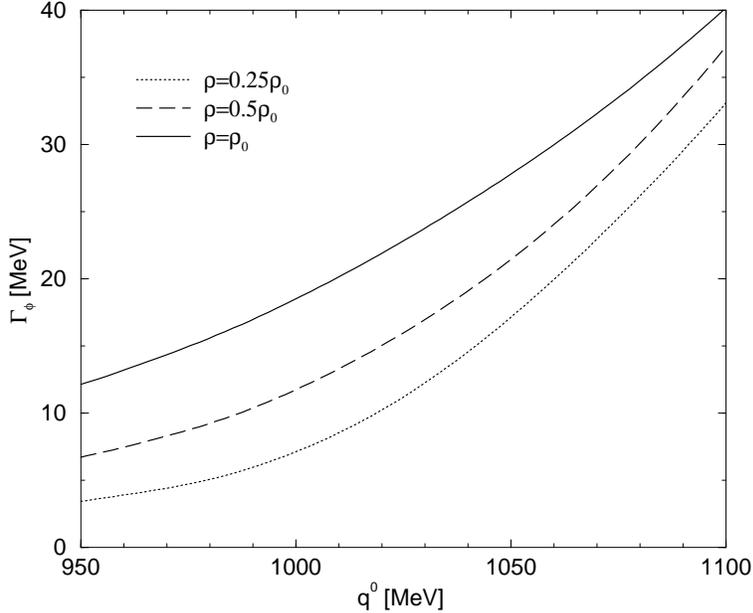}
      \caption{\small Width of the $\phi$ meson in nuclear matter at
three densities ($\rho=0.25\rho_0,0.5\rho_0,\rho_0$) as a function of the
$\phi$ energy.}
        \label{fig:phifig4}
\end{figure}

In  Fig.~\ref{fig:phifig4} we show the total width of the $\phi$ as a
function of the $\phi$ energy for
three different densities. As one can see from the figure, the medium effects
are not proportional to the density. The observed density dependence also
differs for the
different energies of the $\phi$.
We have checked that this highly nonlinear density behavior of the
width comes mainly from the s-wave part of the
$\bar{K}$ selfenergy, as already
noted in Ref.~\cite{klilett}. At $\rho=\rho_0$ this s-wave contribution 
to the width is reduced 
by a factor 2.5 with respect to a linear extrapolation from low densities.
The p-wave contribution is more moderately density dependent, but
there is still a reduction of about 30\% at $\rho=\rho_0$ with
respect to the linear extrapolation from low densities, as also
estimated in Ref.~\cite{klilett}. In the present work we
take into account the full density dependence
of the p-wave contribution, which in Refs.~\cite{klinuc,klilett}
was only considered at the
lowest order in the density.

  The analysis done here about the density dependence and the 
p-wave contributions can be used to support the arguments given at the
end of Sect. 2 justifying the use of the
prescription of eq.~(\ref{eq:vertex}) 
to incorporate
the p-wave vertex corrections. We have carried out a
different calculation in which $\vec{q}\,^2$ is substituted in the $\phi$
vertices by the term on the right in expression (\ref{eq:vertex})
divided
by $\widetilde{\Pi}(q)$.  This new prescription is equivalent to the one used
before in the case of just one particle-hole excitation in 
the p-wave part (it
should not be used for the s-wave part where there are no vertex
corrections). However, this new prescription does not modify the p-wave
selfenergy at higher orders in the density, a feature we would like to have 
in our
calculation.  The differences between the two prescriptions are of the
order of 2\% at $\rho=\rho_0$, which gives us an idea of the accuracy of
the calculation.

The width obtained in the present work is about a factor of two smaller than
that obtained in Refs.~\cite{klinuc,klilett}. Had we used 
$f=f_\pi$ the width would have raised only up to 25 MeV.
One might think that
the differences with the result in Ref.~\cite{klilett} should
be attributed to the fact that the latter work
relied on a $\bar{K}$ selfenergy that was not
selfconsistently evaluated \cite{wolfram}. Their in-medium ${\bar K}N$
amplitude included basically the Pauli-blocking effects on the
nucleons in the intermediate ${\bar K}N$ loops. As a result, the ${\bar K}$
strength at $\rho_0$ presents a quite narrow peak \cite{wolfram}, located
100 MeV below the kaon mass. In contrast, 
when the attractive and complex $\bar{K}$
selfenergy is incorporated selfconsistently in the
in-medium scattering equation,
as done in Refs.~\cite{lutz,angels}, 
the peak of the $\bar{K}$ spectral function moves to energies closer to
the free $\bar{K}$ mass, which should result
in a drastic reduction of the $\phi$
width. However, when we use the
${\bar K}$ selfenergy which only incorporates the Pauli-blocking medium
effects,
an approximation which is also discussed in Ref.~\cite{angels},
the width is only 7\%  larger than that obtained using the
selfconsistent ${\bar K}$ dressing. 
The reason is that the extra widening of the selfconsistent
${\bar K}$ spectral function
weakens the reduction caused by the larger mass of the
$\bar{K}$ in the medium. 
Another factor which contributes to obtaining a smaller result than
that found in Refs.~\cite{klinuc,klilett} is the inclusion in the
present work of 
the p-wave $\bar{K}$ selfenergy to all orders in the $\bar{K}$
propagator, while only the lowest order term in the density was
considered in Refs.~\cite{klinuc,klilett}.

The tests done in this section give us an idea of the theoretical uncertainties
of the present model,
which we can estimate in a band between $22-28$ MeV for the $\phi$ width at
normal nuclear matter density. Yet, when these numbers are compared with the
free width of 4.4 MeV, the important message,
shared with Ref.~\cite{klilett},
is that the width of the $\phi$ is appreciably enhanced in nuclei by
nearly an order of magnitude. 

\begin{figure}[htb]
     \centering
      \epsfxsize = 8cm
      \epsfbox{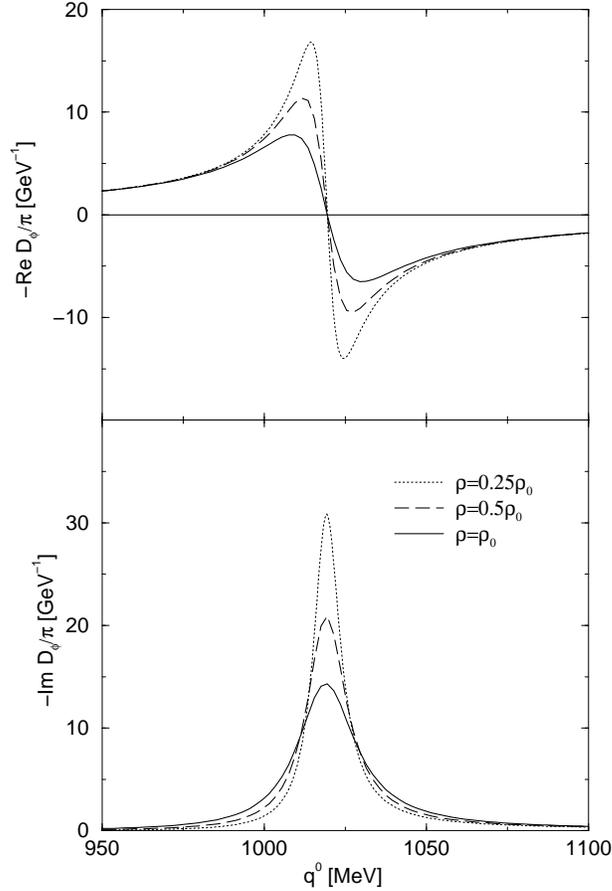}
      \caption{\small Real and Imaginary parts of the $\phi$
meson propagator in nuclear matter at
three densities ($\rho=0.25\rho_0,0.5\rho_0,\rho_0$) as functions of the
$\phi$ energy.}
        \label{fig:phifig5}
\end{figure}

Finally, the real and imaginary parts of the $\phi$ propagator, normalized such that
the later gives the $\phi$ spectral function, are shown in
Fig.~\ref{fig:phifig5} as functions of the $\phi$ energy
for three different densities. 
The mass
of the $\phi$ meson in the medium has been taken to be the same as the
free one,
an assumption that relies 
upon the results of Ref.~\cite{klinuc} where only a small change of
about 1\%  was found.  The novel elements introduced here
in the kaon spectral function should not change this result
qualitatively.
   
The $\phi$ spectral
function shown in the lower pannel of Fig.~\ref{fig:phifig5} gives us an idea
of what one might expect for, let us say,
the invariant mass of $K \bar{K}$ distributions in nuclei around the $\phi$
mass in $\phi$ production experiments, like $\gamma$-nucleus or 
$\pi$-nucleus collisions. These reactions  could be easily done in 
experimental facilities 
like Spring8/Osaka 
or GSI, where programs to produce $\phi$ mesons in the elementary
reactions are
already considered.  Certainly, the results obtained here, extrapolated if
needed at higher densities, should also be of much use in analyses  of
dilepton production in heavy ion collisions around the $\phi$ energy,
which is actually one of the regions studied experimentally.

\section{Conclusions}

  We have studied the different mechanisms of $\phi$ decay in a nuclear
medium. We rely upon the OZI rule which makes the direct coupling of the
$\phi$ to nucleons or nonstrange mesons small, and thus the $\phi$ decay
is still driven by the $K \bar{K}$ or related channels in the medium.
The $\phi$ width in the medium is hence obtained from
previously developed models for the interaction of
kaons in nuclei, which have been tested in $K^-$ atoms recently. The gauge
nature of the $\phi$ as a vector field also introduces some vertex
corrections which lead to a moderate increase of the $\phi$ decay width. 

We have
evaluated this width as a function of the energy of the $\phi$ and the
nuclear density of the medium and have separated the contribution of the
different reaction channels, which can eventually be investigated in some
experiments for $\phi$ production in nuclei.  The relevance of these
findings for the dressing of kaons in nuclei and its relationship to the
eventual kaon condensation in dense matter, as well as the implications
for the
running experiments of dilepton production in heavy ion collisions, 
should
stimulate the performance of $\phi$ production in nuclei with some
particle nucleus collisions which could serve us to test the ideas and
results obtained in the present work. 

\section{Acknowledgements}

 We would like to thank B. Friman for discussions . This work has been
partially supported by the DGICYT contract numbers PB98-1247 and
PB96-0753, and by the EEC-TMR Program, EURODAPHNE, under contract number
CT98-0169.



\begin{thebibliography}{99}
\bibitem{beng} M. Herrmann, B.L. Friman and W. Norenberg, Nucl. Phys. A560
(1993) 411
\bibitem{chanfray} G. Chanfray and P. Schuck, Nucl. Phys. A555 (1993) 329
\bibitem{johan} R. Rapp, G. Chanfray and J. Wambach, Nucl. Phys. A617 
(1997) 472
\bibitem{klinuc} F. Klingl, N. Kaiser and W. Weise, Nucl. Phys. A624 (1997)
527
\bibitem{hades} J. Friese, Prog. Part. Nucl. Phys. 42 (1999) 135
\bibitem{ceres} G. Agakishiev et al., Phys. Lett. B422 (1998) 405
\bibitem{klilett} F. Klingl, T. Waas and W. Weise, Phys. Lett. B431 (1998)
254 
\bibitem{koch} V. Koch, Phys. Lett. B337 (1994) 7
\bibitem{wolfram} T. Waas and W. Weise, Nucl. Phys. A625 (1997) 287
\bibitem{lutz} M. Lutz, Phys. Lett. B426 (1998) 12
\bibitem{angels} A. Ramos and E. Oset, Nucl. Phys. A,  in print.
nucl-th/9906016
\bibitem{kaplan} D.B. Kaplan and A.E. Nelson,  Phys. Lett. B175 (1986) 57
\bibitem{batty} C.J. Batty, Nucl. Phys. A372 (1981) 418
\bibitem{gal} C.J. Batty, E. Friedman and A. Gal, Phys. Rep. 287 (1997) 385
\bibitem{nieves} A. Baca, C. Garc\'{\i}a-Recio and J. Nieves, Nucl. Phys.
A, in print. nucl-th/0001060
\bibitem{satoru} S. Hirenzaki, Y. Okumura, H. Toki, E. Oset and A. Ramos,
Phys. Rev. C61 (2000) 055205
\bibitem{ecker} G. Ecker, J. Gasser, A. Pich and E. de Rafael, Nucl. Phys.
B321 (1989) 311 
\bibitem{pdg} 
C. Caso et al., The European Physical Journal C3 (1998) 1
\bibitem{siegel} N. Kaiser, P.B. Siegel, and W. Weise, Nucl. Phys. A594
(1995) 325; N. Kaiser, T. Waas and W. Weise, Nucl. Phys. A612 (1997) 297
\bibitem{knangels} E. Oset and A. Ramos, Nucl. Phys. A635 (1998) 99
\bibitem{dalitz} R.H. Dalitz and S.F. Tuan, Ann. Phys. (N.Y.) 10 (1960) 307;
R.H. Dalitz, T.C. Wong and G. Rajasekaran, Phys. Rev. 153 (1967) 1617
\bibitem{martin} A.D. Martin, Nucl. Phys. B179 (1981) 33
\bibitem{siegel2} P.B. Siegel and B. Saghai, Phys. Rev. C52 (1995) 392
\bibitem{report} J.A. Oller, E. Oset and A. Ramos, Prog. Part. Nucl.
Phys. 45, in print. hep-th/0002193 
\bibitem{caro} J. Caro, N. Kaiser, S. Wetzel and W. Weise, Nucl. Phys. A
in print, nucl-th/9912035	
\end{thebibliography}
\end{document}